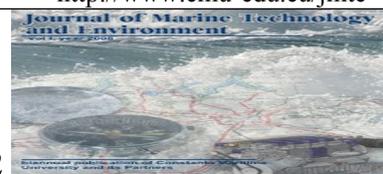

*Journal of Marine technology and Environment*   Year 2018, Vol.2

# R SCRIPTING LIBRARIES FOR COMPARATIVE ANALYSIS OF THE CORRELATION METHODS TO IDENTIFY FACTORS AFFECTING MARIANA TRENCH FORMATION

**Polina Lemenkova** [1]

[1]Ocean University of China, College of Marine Geoscience, 238 Songling Road, Laoshan, 266100, Qingdao, China,
e-mail address: pauline.lemenkova@gmail.com

*Abstract:* Mariana trench is the deepest place on the Earth. It crosses four tectonic plates of the Pacific Ocean: Mariana, Caroline, Pacific and Philippine. The formation of the trench is caused by the complex interconnection of various environmental factors. The aim of this study was to describe and characterize various impact factors affecting formation of the Mariana trench geomorphology and continental margin environments using R programming language and mathematical algorithms of correlation methods written on R code. To record the system of geological, tectonic, geographic, oceanological and bathymetric features affecting Mariana trench, a combination of statistical methods, GIS and R programming codes were applied. The questions answered are as follows: which factors are the most influencing for the Mariana trench morphology, and to what extend do they affect its development? Is sedimental thickness of the ocean trench basement more important factors for the trench formation comparing to the steepness slope angle and aspect degree? Three methods of computing were tested: Pearson correlation, Spearman correlation, Kendall correlation, numerical correlogram, correlation matrix and cross-correlatios to analyze environmental impact factors. The correlogram matrices are computed and visualized by R scripting libraries. Complex usage of programming tools, mathematical statistics and geospatial analysis enabled to get a differentiated understandings of the hadal environments of the Mariana trench. The results revealed following three types of factors having the highest score: geometric (tg° slope angle), geologic (sedimental thickness) and tectonic structure. The results furthermore indicated that tectonic plates, sedimental thickness of the trench basement and igneous volcanic areas causing earthquakes play the most essential role in the geomorphology of the trench.

## 1. INTRODUCTION

Mariana trench is a long and narrow topographic depression of the sea floor in the west Pacific ocean, 200 km to the east of the Mariana Islands, located east of the Philippines. It is the deepest part of the ocean with a maximal depth of $10,984 \pm 25$ m (95%) at 11.329903°N / 142.199305°E m at the Challenger Deep [6].

Mariana trench crosses four tectonic plates: Mariana, Caroline, Philippine and Pacific. It has a distinctive morphological feature of the four convergent Pacific plate boundaries, along which lithospheric plates move towards each other. The total length of the trench measures about 2,550 km long approximately.

Various environmental factors affect the geomorphological structure, formation and development of the trench, the most important of which include the following: 1) tectonics: slabs and tectonics plates; 2) bathymetric features, depth values; 3) geographic location affecting slope aspect degree; 4) geologic structure of the underlying basement and sedimental thickness of the bottom layer. Briefly describe the most important factors below.

## 2. DESCRIPTIONS OF MARIANA TRENCH

Mariana trench presents a strongly elongated, arched in plan and lesser rectilinear depression stretching for hundreds kilometers and having a narrow depression on the ocean floor with very steep slopes and the depths of 3-5 km width of the bottoms rarely exceeding 5 km in breadth.

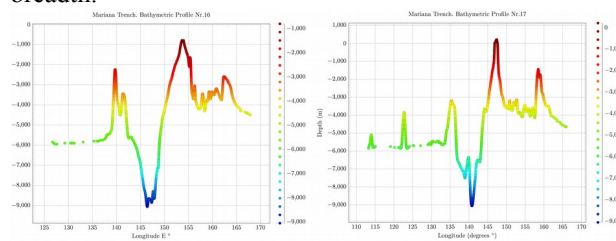

Figure 1. Selected bathymetric profiles of the Mariana trench. Plot visualization: LaTeX.

Its transverse profile is strongly asymmetric: the slopes of the Mariana trench are higher on the side of the island arc. The slopes of the trench are dissected by deep underwater canyons (Fig.1). On the slopes of the trench are also often found various narrow steps. Mariana trench has complicated steps of various shapes and sizes, caused by active tectonic and sedimental processes. The steepness of its geomorphic depth averages in 4-5 degrees (Fig. 2), but its individual parts can be limited to steeper slopes as subjects to the gravitational flow system of the submarine canyons and valleys.

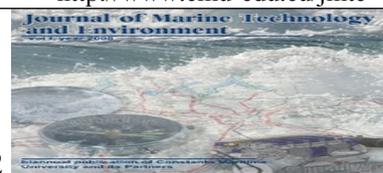

Mariana trench is the largest structural trap located in the continental margins of the Pacific: the sediments are being carrying by the ocean waves in a clockwise direction, passing through the trenches on the west of the Pacific (i.e. the Kermadec Trench and the Tonga Trench. Through the Samoan Passage). They furthermore flow northwest across the equator to the east Mariana.

## 2.1    Submarine earthquakes

Deep earthquakes are important factors causing system of cracks of the Mariana trench. The occurrence of the deep earthquakes could be explained by various factors. For instance, transformation faulting was proposed as the mechanism for deep earthquakes by [14]. This research is largely based on the basis of experimental observations [14]; [7]; [13]; [4] and involves a shear instability that develops during the incipient transformation of metastable olivine to wadsleyite or ringwoodite.

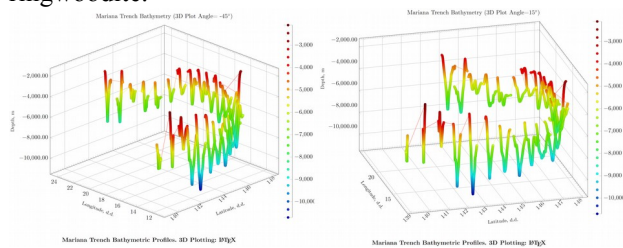

Figure 2. 3D-view of the 25 bathymetric profiles of the Mariana trench. Plot visualization: LaTeX.

The horizontal tensional stress during the earthquake shaking triggers development of the system of cracks in the Mariana trench, because the stress difference under static conditions is caused by the slope inclination. The majority of the cracks within the Mariana trench are open tensile failures and are situated in areas of very gentle slope or horizontal parts of the seabed, in most cases near the edges of steep cliffs which delimit the terraces. There is no lateral nor vertical dislocations on both sides of the cracks, so they are open fractures.

## 2.2    Tectonic plates subduction

Many physical and chemical processes interact during tectonic plates subduction, and the complexity of the system necessitates an interdisciplinary approach involving seismology, mineral physics, geochemistry, petrology, structural geology, rock mechanics, and geodynamic modeling. For example, phase transformation kinetics determine buoyancy, rates of subduction and, therefore, thermal structure, with the latter feeding back to affect the kinetics. Rheology and therefore large scale slab dynamics may also be affected by mineral transformations and their kinetics.

The rheology of mantle minerals, though having considerable importance in controlling the subduction process, is exceedingly difficult to study experimentally, due to the experimental limitations, since rheology at pressures of the transition zone and lower mantle [12]. An increase of intrinsic density or viscosity with depth as well as phase transitions with a negative Clapeyron slope can all inhibit or delay deep subduction, and the tectonic conditions and the relative motion of the tectonic plates at Earth's surface also play an important role. The effects of subduction of crustal rocks during continental collision consists in following subduction to depths greater than 90km. Slices of continental crust have been exhumed back to the surface very rapidly, thus ensuring the survival of the high-pressure minerals.

## 2.3    Tectonic slab dynamics

Slab dynamics if one of the important drivers for the trench formation. Effects of slab mineralogy and phase chemistry on the subduction dynamics (buoyancy, stress field), kinematics (rate of subduction and plate motion), elasticity (deformation and seismic wave speed), thermometry (effects of latent heat, isobaric superheating) and seismicity (due to adiabatic shear instabilities) are well discussed [3].

The morphology and development of the subducted slabs are more complex than modeled [5], on either whole mantle flow or convective layering at 660 km depth. Apparently, slabs can deflect horizontally in the upper mantle transition zone beneath some convergent margins whereas penetration to lower mantle depths can occur beneath other island arc segments. The different styles of the subduction across the upper mantle transition zone and trench formation.

Not all seismic zones in subducted slabs of the Mariana trench are continuous and that some deep events seem to occur isolated from those at shallower depths. Thus, it is noticed [10] that the continuity of the slab across gaps in the Wadati–Benioff seismic zone between deep earthquake clusters and shallow seismicity and they argue that only the deep earthquakes beneath the north Fiji basin and the deep events beneath New Zealand may occur as detached events with no mechanical connection to the surface. Using the observation and triggering mechanism of high-frequency oceanic "T" waves, [2], it has been argued [8], [9] for the mechanical continuity of most slabs across the deep seismicity gap.

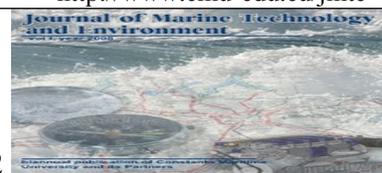

Journal of Marine technology and Environment    Year 2018, Vol.2

## 2.4    Deep seismicity

Based on the correlations between strength and temperature, the deep seismicity is likely to result from plastic instabilities, with the distribution of earthquakes being related to strength distribution and therefore slab mineralogy. A complex rheological structure for subducting slabs has been proposed [12], due to the effects of grain size reduction during phase transformations. Based on their model, rapidly subducting and therefore relatively cold slabs should be weaker than slowly subducting warm slabs. Not all seismic zones in subducted slabs are continuous and that some deep events seem to occur isolated from those at shallower depths [10].

As for the structure, magnetism, and dynamics of subduction zones, the continuation of the low wave speed, high attenuation (low Q), and seismically anisotropic wedge are normally near 400 km depth beneath some back arcs. This may be explained by the persistence of hydrous phases to that depth and that magnetic systems are not limited to near-surface regions of the mantle. The correlation also exist between the dehydration of the slab, rupture nucleation and crustal earthquake triggering.

## 2.5    Seafloor spreading

Processes of the deep-sea terrigenous sedimentation are formed by the transfer of erosion materials from the adjacent land. The main processes are: transportation, deposition and re-deposition of the sedimentation materials. There are two general types of sedimentation in the Mariana trench.

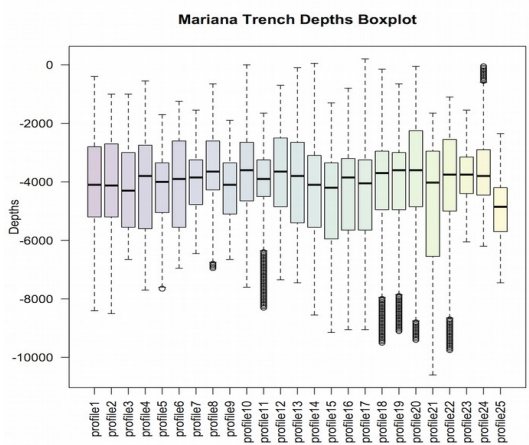

Figure 3. Box plot histograms on data distribution, Mariana trench bathymetric profiles 1:25.

The first one is represented by the pelagic sediments, accumulated as a result of gravitational forces moving suspended matter from the water column into the deeper parts. The second type is represented by the aleurite-clay sediments. In high latitudes they may be mixed with the iceberg sediments. The speed of such terrigenous sedimentation can reach 175-200 mm/thousand years. The most interest have sediments in the bottoms of deep ocean trenches, that mainly accumulate horizontally sand and silt turbidites with gradational stratification.

## 3.    METHODS

### 3.1.    Analysis of bathymetric data distribution by statistical box plots

The created box plot of the depths distribution show summary statistics, that is range and quartiles of the observation points across the bathymetric profiles. A box plot (Fig.3) graphically displays the frequencies of a bathymetric observations data set of the 25 profiles of Mariana trench. It plot the frequency or count, on the depth y-axis (vertical) and the variable being measured on the x-axis, that is profiles (horizontal). The box plot presented in this research (Fig.3) was created by the default R library {stat} graphically representing five most important descriptive values for a bathymetric data set.

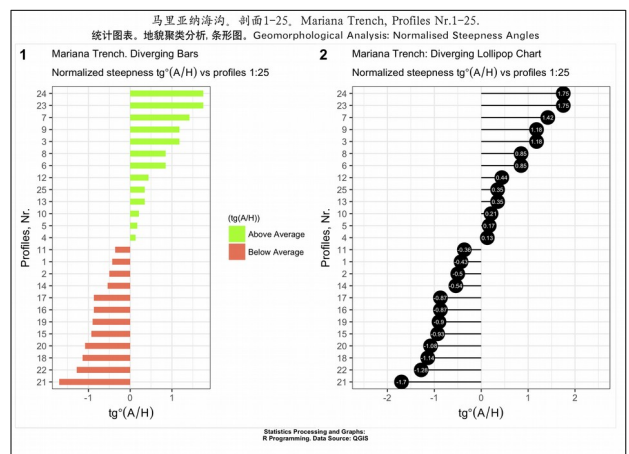

Figure 4. Normalized steepness slope angle, bathymetric profiles 1:25 of Mariana trench.

These values include the following bathymetric values: minimum value, the first quartile (0.25), the median, the third quartile (0.75), and the maximum value (that is, absolute depth). When graphing this five-number summary, only the horizontal axis displays values, while a vertical line is placed above each of the summary numbers showing depth values. A box is drawn around the middle three lines (first quartile, median, and third quartile) and two lines are drawn from the box's edges to the two endpoints (minimum and maximum values of depths).

### 3.2    Calculation of the normalized steepness angle of the Mariana trench

http://www.cmu-edu.eu/jmte

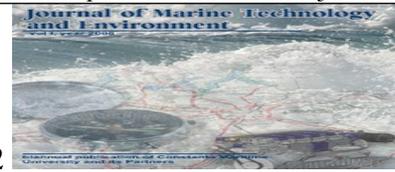

Journal of Marine technology and Environment   Year 2018, Vol.2

The normalized steepness was computed as a steepness of dominance hierarchies of slope angle (Fig.4). Steepness is defined as the absolute slope of the straight line fitted to the normalized scores. The normalized scores were obtained on the basis of dominance indices corrected for chance or by means of proportions of wins. Given an observed data set, it computes hierarchy's steepness and estimates statistical significance by means of a randomization test. The following R programming code was used for execution:

```
MDF$"profile name" <- rownames(MDF)  # create new
column for car names
 # step-1. Re-calculate argument values (X-axis) into
normalized through the difference between mean and
standard deviation
MDF$norm_tg_angle <- round((MDF$tg_angle –
mean(MDF$tg_angle))/sd(MDF$tg_angle), 2)  #
compute normalized tg_angle
 # step-2. Distribute values of the normalized argument
into "above" and "below" mean
MDF$angle_type <- ifelse(MDF$norm_tg_angle < 0,
"below", "above")  # above / below avg flag
 # step-3. Sort dataframe
MDF <- MDF[order(MDF$norm_tg_angle), ]  # sort
 # step-4. Values Y (here: profile names) convert them
into factor names
MDF$"profile name" <- factor(MDF$"profile name",
levels = MDF$"profile name")  # convert to factor to
retain sorted order in plot
class(MDF$profile name) # check up class
# [1] "factor"
MDF # take a look at new dataframe
 # draw 2 plots using dataframe MDF
 # step-5. Diverging bars plot
Diverging_Bars<- ggplot(MDF, aes(x = MDF$"profile
name", y = MDF$norm_tg_angle, label =
MDF$norm_tg_angle)) +
 geom_bar(stat='identity', aes(fill = MDF$angle_type),
width=.5) +
 xlab("Profiles, Nr.") +
 ylab(expression(tg*degree*(A/H))) +
 scale_fill_manual(name="(tg(A/H))",
              labels = c("Above Average", "Below
Average"),
              values = c("above"="lawngreen",
"below"="coral1")) +
 labs(title= "Mariana Trench. Diverging Bars",
       subtitle=expression(paste("Normalized steepness
", tg*degree*(A/H), " vs profiles 1:25"))) +
 coord_flip() +
 theme(plot.title = element_text(size = 10),
            legend.title = element_text(size=8),
 legend.text = element_text(colour="black", size = 8))
Diverging_Bars
 # step-6. Plotting chart
```

```
Chart <- ggplot(MDF, aes(x = MDF$"profile name", y =
MDF$norm_tg_angle, label = MDF$norm_tg_angle)) +
 xlab("Profiles, Nr.") +
 ylab(expression(tg*degree*(A/H))) +
 geom_point(stat='identity', fill="black", size=6)  +
 geom_segment(aes(y = 0, x = MDF$"profile name",
 yend = MDF$norm_tg_angle, xend = MDF$"profile
name"), color = "black") +
 geom_text(color="white", size=2) +
 labs(title="Mariana Trench: Diverging Lollipop Chart",
 subtitle=expression(paste("Normalized steepness ",
 tg*degree*(A/H), " vs profiles 1:25"))) +
 ylim(-2.5, 2.5) +
coord_flip() +
  theme(plot.title = element_text(size = 10), legend.title
= element_text(size=8), legend.text =
 element_text(colour="black", size = 8))
 # step-7. Plot both charts together as one figure
figure <-plot_grid(Diverging_Bars, Lollipop, labels =
c("1", "2"), ncol = 2, nrow = 1)
```

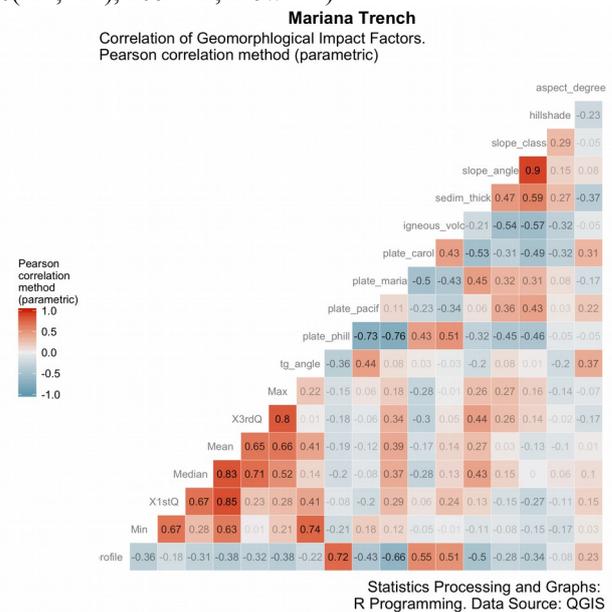

Figure 5. Visualized Pearson correlation for Mariana trench geomorphological impact factors

### 3.3    Correlation analysis of impact factors

Correlation coefficient is a numerical measure of direction and strength of linear correlation between various environmental variables, i.e.  how strong a relationship is between geomorphological, bathymetric and geological variables is across four different tectonic plates.

### 3.3.1.  Computing Pearson correlation

Developed by Karl Pearson from a related idea of Francis Galton in the 1880s, a Pearson product moment correlation coefficient, also know as the bivariate correlation, is a measure of the linear correlation between two variables:

http://www.cmu-edu.eu/jmte

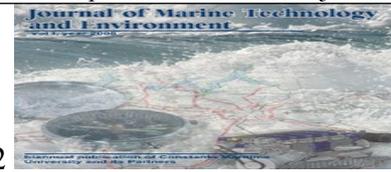

Journal of Marine technology and Environment   Year 2018, Vol.2

$$r = \sum (x - mx)(y - my)/\sqrt{\sum (x - mx)2\sum y - n}$$

(1)

In this case the environmental variables have been tested by the Pearson correlation: sedimental thickness, slope angle, slope class, aspect degree, hill shade in values, location of igneous volcanic zones across the abyssal valley, tectonics by four plates (Mariana, Caroline, Philippine and Pacific), a bunch of bathymetric values, that is maximal and minila depth values, median means value for each profile, tangens angle degree.

Pearson correlation has a value between +1 and −1, where 1 is total positive linear correlation, 0 is no linear correlation, and −1 is total negative linear correlation. It has been presented on Fig. 5. It was drawn by calling following R programming script:

```
MDF <- read.csv("Morphology.csv", header=TRUE, sep = ",")
MDF <- na.omit(MDF)
row.has.na <- apply(MDF, 1, function(x){any(is.na(x))})
sum(row.has.na)
head(MDF)
# Check correlation between variables
cor(MDF)
# Visualization of correlations
# Pearson correlation coefficients, using pairwise
 observations
gp<- ggcorr(data=MDF, method = c("everything",
 "pearson"),
 name = "\nPearson \ncorrelation \nmethod \
n(parametric)",
 label = TRUE, label_size = 2, label_round = 2,
 label_alpha = TRUE,
 hjust = 0.75, size = 3, color = "grey50", legend.position
 = "left")
gpt<- gp + labs(title="Mariana Trench",
 subtitle = "Correlation of Geomorphlogical Impact
 Factors \nPearson correlation method (parametric)",
 caption = "Statistics Processing and Graphs: \nR
 Programming. Data Source: QGIS") +
theme(plot.title = element_text(family = "Times New
Roman", face = 2, size = 12),
        plot.subtitle = element_text(family = "Times New
Roman", face = 1, size = 10),
        plot.caption = element_text(family = "Times New
Roman", face = 2, size = 8))
```

In this case the environmental variables have been tested by the Pearson correlation: sedimental thickness, slope angle, slope class, aspect degree, hill shade in values, location of igneous volcanic zones across the abyssal valley, tectonics by four plates (Mariana, Caroline, Philippine and Pacific), a bunch of bathymetric values, that is maximal and minila depth values, median means value for each profile, tangens angle degree.

### 3.3.2. Computing Spearman correlation

Named after Charles Spearman, a Spearman non-parametric coefficient is handling ties between the variables less properly comparing to previously computed (see 2.6.1) Pearson coefficient (Fig.6). The Spearman's has been computed on sets of paired rankings between the environmental factors. On the Fig.6 the more correlated variables are represented by the colors of green. The size of the circle also depends on the correlation value.

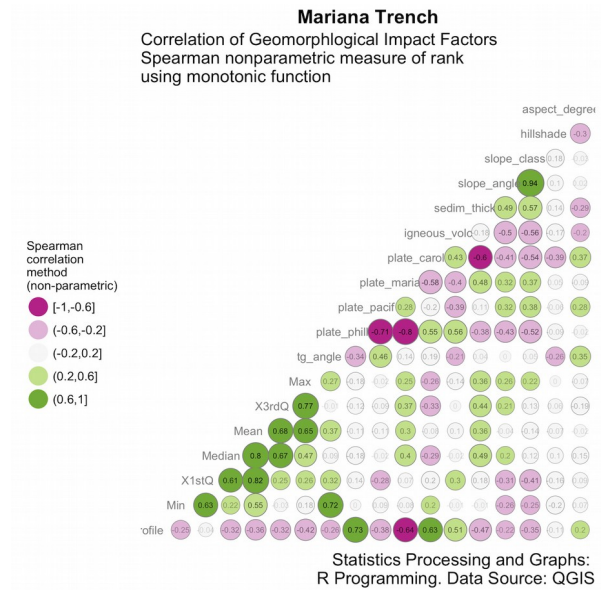

Figure 6. Visualized Spearman correlation ranking for Mariana trench geomorphological impact factors

Spearman's rank correlation coefficient is denoted by $r_s$. It is given by the following formula:

$$r_s = 1 - (6\sum d_i 2)/(n(n2 - 1))$$

(2)

Here $d_i$ represents the difference in the ranks given to the values of the variable for each item of the particular data The formula of Spearman's rank correlation coefficient (2) is applied in cases when there are no tied ranks. Technically, the R programming script for Pearson correlation was changed in case of Spearman by adding following code:

```
gs<-ggcorr(data=MDF,   method = c("everything",
"spearman"), geom = "circle", nbreaks = 5,
 min_size = 3, max_size = 9, palette = "PiYG",
 name = "\nSpearman \ncorrelation \nmethod \n(non-
parametric)",
 label = TRUE, label_size = 2, label_round = 2,
 label_alpha = TRUE,
```

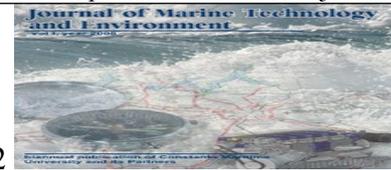

hjust = 0.75, size = 3, color = "grey50", legend.position = "left")

Two plots were then plotted together by calling:figure <- plot_grid(gpt, gst, labels = c("1", "2"), ncol = 2, nrow = 1)

On Fig. 6 the measure of rank correlation, that is a statistical dependence between the rankings of two environmental variables has been computed using {GGalt} library.

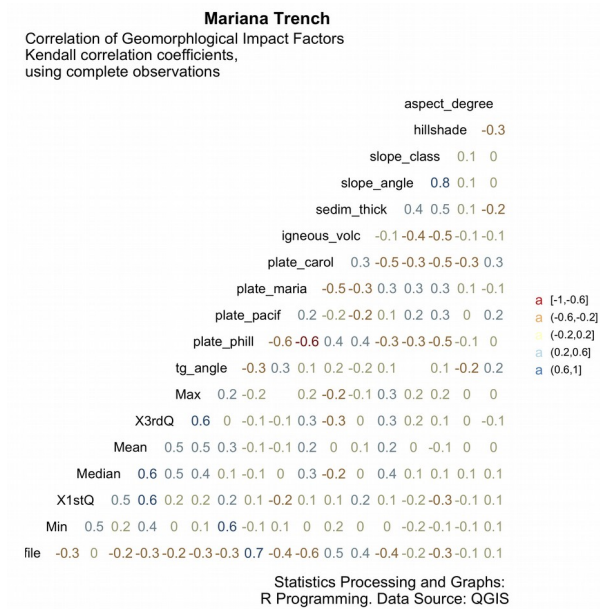

Figure 7. Visualized Kendall tau-correlation for Mariana trench geomorphological impact factors

Spearman correlation shows (Fig.6) how well the relationship between every pair of two environmental variables can be described using a monotonic function and comparing following factors: tectonics, geology, geographic, bathymetric and magmatism. In other words, Spearman's correlation assesses monotonic relationships, whether linear or not between the bunch of available factors. If there are no repeated data values, a perfect Spearman correlation of +1 or −1 occurs when each of the variables is a perfect monotone function of the other.

### 3.3.3. Computing Kendall correlation

Alike to Spearman coefficient, Kendall tau rank correlation, invented by Maurice Kendall, is also a non-parametric test for statistical dependence between environmental ordinal (or rank-transformed) variables.

Kendall correlation distance is defined as follow:

$$\tau = nc - nd \, 1/2 \, n(n-1) \qquad (3)$$

Kendall rank correlation tau coefficient is a statistically used to measure the ordinal association between two measured quantities, that is environmental factors in this case. Based on the tau coefficient test, a non-parametric hypothesis test for statistical dependence has been computed (Fig.7).

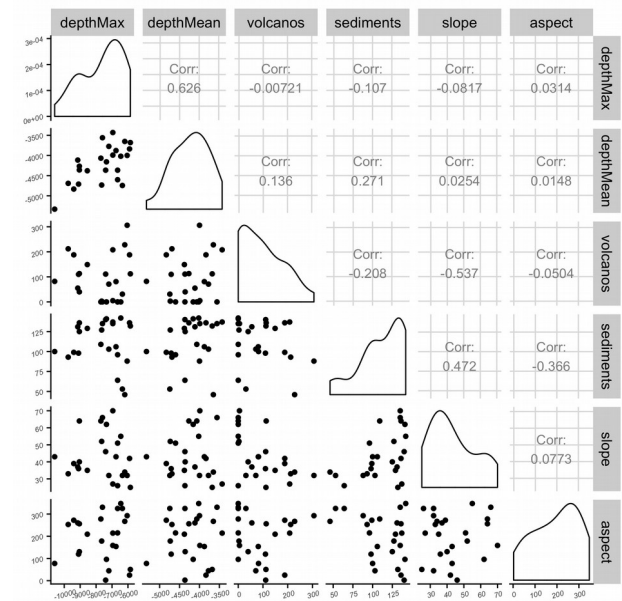

Figure 8. Numerical scatterplot matrix of correlation of the geomorphological impact factors of Mariana trench

It measures a rank correlation: the similarity of the orderings of the data when ranked by each of the quantities. Technically, the script for computing Lendall correlation is as follows:

gk<- ggcorr(data=MDF, method = c("complete", "kendall"),
  geom = "text", nbreaks = 5, palette = "RdYlBu", hjust = 1, label = TRUE, label_alpha = 0.4)

gkt<- gk + labs(title="Mariana Trench", subtitle = "Correlation of Geomorphlogical Impact Factors \nKendall correlation coefficients, \nusing complete observations",
    caption = "Statistics Processing and Graphs: \nR Programming. Data Source: QGIS")

Similar to Spearman coefficient, but unlike Spearman's, can handle ties. Moreover, there are three Kendall tau statistics: tau-a, tau-b, and tau-c, of which tau-b is specifically adapted to handle ties. The tau-b statistic handles ties, i.e., each pair of members of the environmental variables, e.g. tectonics and geological factors, or geographic location and magmatism, or sedimental thickness layer versus closeness of the igneous volcanic spots, − will have the same ordinal value by a divisor term. The latter represents the geometric mean between the number of environmental factor pairs that are not tied on factor one and the number not tied on two.

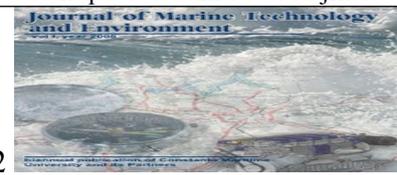

Numerical scatterplot matrix of correlation of the geomorphological impact factors of Mariana trench (Fig. 8) and cross-correlation matrix of the depth values in bathymetric profiles of the Mariana trench (Fig.9) are drawn using R scripting libraries.

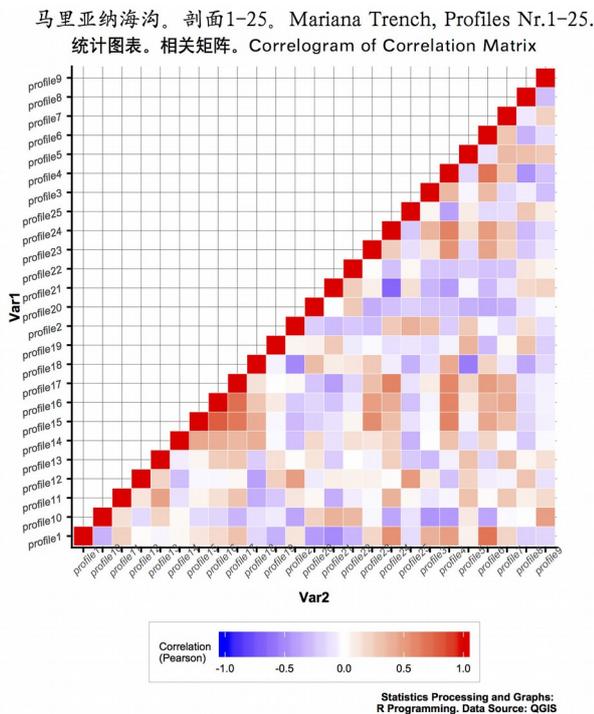

Figure 9. Cross-correlation matrix of the depth values in bathymetric profiles of the Mariana trench

### 3.4. Computing principal component analysis

The Principal Component Analysis (Fig. 10) enabled to visualize eigenvectors as showing major direction and vector length for the principal components affecting the categorical values. Through the PCA a statistical procedure using an orthogonal transformation to convert a set of bathymetric depth observations of possibly correlated variables, has been performed. Thus, the direction of the eigenvectors shows the depth of the Mariana trench has been highly influenced by its geomorphic settings, particularly its location, by 25 profiles, as well as the similarities between the profiles.

### 4. RESULTS

Current studies reveled that there are factors influencing Mariana trench geomorphic structure the most. The most affecting factors are as follows: sedimental thickness of the basement, slope angle steepness degree, angle aspect, bathymetric factors, such as depth at basement, means, median and minimal values, closeness of the igneous volcanic areas causing possible earthquakes, and geographic location across four tectonic plates – Mariana, Pacific Philippine and Caroline.

Summary of the results is being represented on the Fig. 11 as a hierarchical tree map of impact factors. Mariana trench is an important integral feature of the active continental margins of west Pacific Ocean. Mainland and oceanic sides of the trench have complicated steps of various shapes and sizes, caused by active tectonic and sedimental processes. The steepness of the trench depth averages in 4-5 degrees.

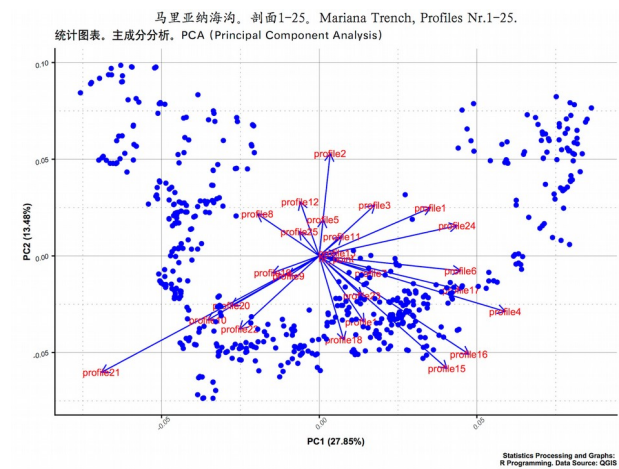

Figure 10. Principal component analysis of correlation of the bathymetric values of profiles of the Mariana trench

Adjacent tectonic plates, however, have more steep angle with an average slopes of 10 to 15 degrees, but their individual parts can be limited to steeper slopes as subjects to the gravitational flow system of the submarine canyons and valleys. Complex distribution of the various geomorphic material on the adjacent abyssal plains of the ocean contributes to the formation of the geomorphic features of the particular region of the ocean bottom in Mariana trench. It was furthermore found that slope degree and amplitude has important impact on the sedimental thickness, while the aspect degree has lesser effect.

### 5. CONCLUSIONS

The presented research work focused on the classification of the influencing environmental factors of three main groups (tectonics, bathymetry, geology) affecting formation of the Mariana trench. The research is expected to lead to a significant improvement of our understanding of the relevant influencing geomorphic factors for the analysis of the ocean trenches.

Additionally, the review of the correlation methods presented în the current research can be applied methodologically for comparing and benchmarking different research schemes on ocean trench analysis.

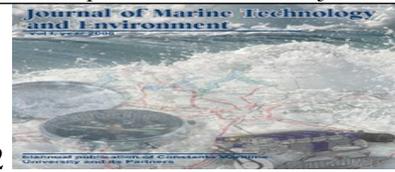

Journal of Marine technology and Environment   Year 2018, Vol.2

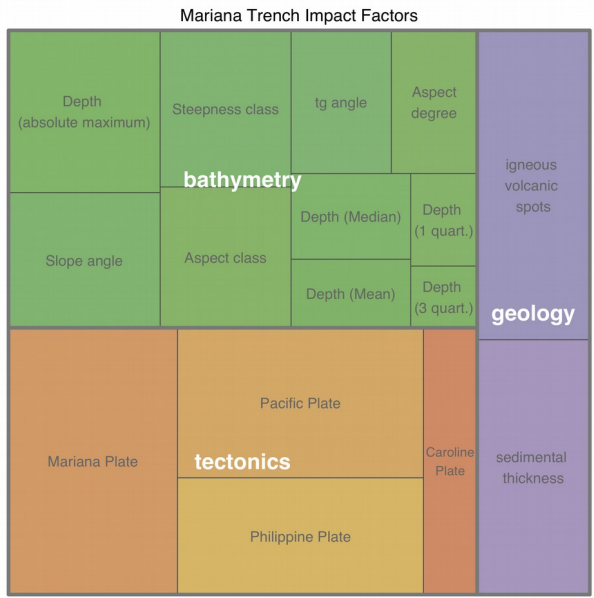

Figure 11. Hierarchical tree map of impact factors affecting formation of the Mariana trench

Furthermore it can be used for analysis of submarine earthquake connections with ocean trench formation and tectonic slab movement (e.g. for the selection of research methods) or the controlling of impact factors during the geospatial analysis.

## 6.     ACKNOWLEDGMENTS

The funding of this research has been provided by the China Scholarship Council (CSC) State Oceanic Administration (SOA), Marine Scholarship of China [Grant # 2016SOA002, 2016].